\documentclass[a4paper,12pt]{article}

\usepackage{latexsym,amsmath,amsfonts,amssymb}
\usepackage{multirow}
\usepackage{float}
\usepackage{epsfig}
\usepackage{feynmp}
\usepackage{graphicx}
\usepackage{subfigure}
\usepackage{mathrsfs,mathtools}
\usepackage{pstricks,pst-node,pst-text,pst-3d}
\usepackage{verbatim}

\newcommand{\ex}[1]{{\rm e}^{#1}} \def\ii{{\rm i}}
\newcommand{\Tr}{{\rm Tr}}
\newcommand{\sect}[1]{\setcounter{equation}{0}\section{#1}}

\begin{document}

\begin{titlepage}

\setcounter{page}{0}

\begin{flushright}
{QMUL-PH-08-18}
\end{flushright}

\vspace{0.6cm}

\begin{center}
{\Large \bf Mass corrections in string theory and lattice field theory.} \\ 

\vskip 0.8cm

{\bf Luigi Del Debbio and Eoin Kerrane}\\
{\sl
SUPA, School of Physics and Astronomy, \\
University of Edinburgh \\
Edinburgh EH9 3JZ, Scotland}

\vskip .3cm

{\bf Rodolfo Russo}\\
{\sl 
Centre for Research in String Theory \\ Department of Physics\\
Queen Mary, University of London\\
Mile End Road, London, E1 4NS,
England}\\

\vskip 1.2cm

\end{center}

\begin{abstract}

  Kaluza-Klein compactifications of higher dimensional Yang--Mills
  theories contain a number of four dimensional scalars corresponding
  to the internal components of the gauge field. While at tree-level
  the scalar zero modes are massless, it is well known that quantum
  corrections make them massive. We compute these radiative
  corrections at 1--loop in an effective field theory framework, using
  the background field method and proper Schwinger--time
  regularization. In order to clarify the proper treatment of the sum
  over KK--modes in the effective field theory approach, we consider
  the same problem in two different UV completions of Yang--Mills:
  string theory and lattice field theory. In both cases, when the
  compactification radius $R$ is much bigger than the scale of the UV
  completion ($R \gg \sqrt{\alpha'},a$), we recover a mass
  renormalization that is independent of the UV scale and agrees with
  the one derived in the effective field theory approach. These
  results support the idea that the value of the mass corrections is,
  in this regime, universal for any UV completion that respects
  locality and gauge invariance. The string analysis suggests that
  this property holds also at higher loops. The lattice analysis
  suggests that the mass of the adjoint scalars appearing in $\mathcal
  N=2,4$ Super Yang--Mills is highly suppressed, even if the lattice
  regularization breaks all supersymmetries explicitly. This is due to
  an interplay between the higher--dimensional gauge invariance and
  the degeneracy of bosonic and fermionic degrees of freedom.

\end{abstract}

\vfill

\end{titlepage}

\sect{Introduction}
\label{sec:intro}

Gauge theories compactified on a circle or a torus appear in various
different physical contexts. For instance, the reduction on a circle
from four to three dimensions is relevant for studying the finite
temperature effects, while toroidal compactification from $D$ to four
dimensions provides the simplest possible toy model for extra
dimensional theories. The components of the gauge field along the
compact dimensions appear as scalars in the effective field theory for
the non--compact space. Imposing periodic boundary conditions, these
scalars contain a massless zero mode. It has been known for a long
time that these massless modes are lifted by radiative
corrections~\cite{Hosotani:1983xw}: in a Yang--Mills theory
compactified on the circle $S^1$ it is possible to write a gauge
invariant mass term and so we expect to find a non-zero 1--loop
correction $\delta m^2$ that vanishes in the limit $R\to\infty$. The
quantum mass corrections to the zero and higher Kaluza--Klein modes
were thoroughly studied in the context of extra--dimensional field
theories~\cite{Hatanaka:1998yp,vonGersdorff:2002as,Cheng:2002iz}, with
particular attention towards phenomenological applications, see for
instance~\cite{Antoniadis:2001cv}.  However this effective field
theory approach has some shortfalls: since a higher dimensional field
theory is non--renormalizable, a sensitivity to the UV physics can
appear which depends on the regularization scheme; moreover it is not
entirely clear how to treat rigorously the sum over the Kaluza--Klein
modes. Different approaches for computing the vacuum polarization have
been proposed and seem to give mostly consistent results, see
e.g. Refs.~\cite{Ghilencea:2001ug,Kim:2001gk,Ghilencea:2001bv,GrootNibbelink:2001bx,Puchwein:2003jq,Falkowski:2003iy,Alvarez:2006we,Alvarez:2006sf,Irges:2007qq,Bauman:2007rt,Bauman:2008rr}. The
finiteness of the scalar mass can also be obtained in the effective
potential approach, as discussed in Ref.~\cite{Hosotani:2005fk}. The
equivalence of the two approaches and a first step towards a two--loop
calculation were presented in Ref.~\cite{Hosotani:2007kn}.

In this paper we want to explore dimensional reduction as a tool for
defining extended supersymmetric theories. In particular we try to
find new hints for defining extended supersymmetry on the lattice
without fine--tuning. In order to obtain some quantitative
information, we concentrate on the 1--loop corrections to the mass of
the adjoint scalar field that is obtained from the zero mode of the
gauge field component along a compactified direction, and try to
disentangle the high--energy (i.e. cut--off scale) contributions from
the low--energy ones. Insight on this problem is obtained by
considering the quantum corrections to the mass of the Kaluza--Klein
zero modes in two different UV completions of Yang--Mills: string
theory and lattice field theory. As we shall see below, both cases are
concrete examples of finite theories, where explicit and unambiguous
computations can be performed. Even though the computational
techniques are different in the two cases, a clear physical picture
emerges from the comparison of the two computations: the leading order
correction to the scalar mass is universal and agrees with the result
obtained from an effective field theory computation.

The string theory computation is most easily compared with the quantum
field theory one if the latter is performed using the background field
method and a Schwinger--time regularization, see
Ref.~\cite{Frizzo:2000ez} and references therein. The setup for the
background field calculation is discussed in
Sect.~\ref{sec:qft}. However our goal in this work is to obtain a
quantitative result for the string and lattice theory computations; we
summarize the effective field theory computation mainly to set-up a
common notation and facilitate the comparison. Our results in
Sect.~\ref{sec:qft} agree with the well--known results in
Refs.~\cite{Hatanaka:1998yp,vonGersdorff:2002as,Cheng:2002iz} for the
mass renormalization of the scalar mode, thereby providing a new
consistency check.  In string theory we follow the procedure outlined
in Refs.~\cite{Weinberg:1985je,Seiberg:1986ea,Sen:1987bd} and relate
the mass shift $\delta m^2$ to the correlator of two vertex
operators. Note that the string computation requires a prescription to
regulate the divergences that appear when the two vertices are very
close on the world-sheet. These divergences are automatically
regulated when the soft insertions of the external states are resummed
and one derives the radiative mass corrections from the effective
action, as done in
Refs.~\cite{Antoniadis:2000tq,Kitazawa:2008tb}. Even if this approach
is very efficient for untwisted string states such ours, it cannot be
applied to the case of twisted states. Thus it is interesting to
follow Refs.~\cite{Weinberg:1985je,Seiberg:1986ea,Sen:1987bd}, as we
do, and extract the mass renormalization from the two--point function
(see Ref.~\cite{Minahan:1989cb} for an application of this approach in
the context of closed string theory). With the proper prescription for
the short--distances divergences on the world--sheet, we verify that
the mass corrections to the components of the gauge field in the
non--compactfied dimensions vanish, as dictated by the
four--dimensional gauge symmetry. Having regulated these divergences,
the string theory techniques are readily extended to the case of
compact space--time dimensions. In close analogy with the non--compact
case, we find that the string calculation is easily mapped into the
quantum field theory calculation and there is quantitative agreement
between the two approaches when the string scale is much lower than
the compactification scale ($\sqrt{\alpha'}\ll R$).

In order to study the theory defined on a discrete space--time
lattice, we generalize the techniques developed in
Ref.~\cite{Kaste:1997ks} in the context of finite--temperature field
theories. Again, when the lattice scale $a$ is much lower than the
compactification scale, the mass generated by radiative corrections
for the component of the gauge field in the compact dimension is found
to be identical to the one obtained in the effective theory
calculation and thus to the string theory one in the regime
$\sqrt{\alpha'}\ll R$. Notice that the lattice and the string theory
calculations deal with the sum over the Kaluza-Klein modes in a very
different way: the string UV completion provides a setup where the
so-called Kaluza-Klein regularization is implemented in a consistent
way and the sums run over {\em all} the modes; on the contrary,
lattice gauge theory provides a gauge invariant way of implementing a
hard cutoff on the integrals and sums and only modes of energies up to
$1/a$ are considered. The fact that these two different approaches
yield the same result in the limit $a,\sqrt{\alpha'}\ll R$ suggests
that all UV completions that respect locality and gauge invariance
yield a leading order contribution to the scalar mass that is
completely captured by an effective field theory approach. The
physical reason is that the high--energy modes in the UV completion
see the extra--dimensions as uncompact and so do not contribute to the
mass renormalization because of the higher--dimensional gauge
symmetry.

We find that a similar pattern holds also for the 1--loop contribution
of fermions in lattice perturbation theory. The fermionic contribution
can actually be written in a form that is very close to the bosonic
one. As a consequence, we find that the leading terms in the bosonic
and fermionic contributions to the mass renormalization of the adjoint
scalar field cancel whenever the number of degrees of freedom are
equal. Hence the mass renormalization of the scalar field is highly
suppressed if a supersymmetric theory is dimensionally reduced. Our
computation provides an explicit one--loop realization of the
mechanism suggested in Ref.~\cite{Kaplan:1999jn}, and supports
the interesting possibility that Yang--Mills theories with
extended supersymmetry can be defined on the lattice without any
fine--tuning by dimensional reduction of a higher dimensional
$\mathcal N=1$ theory, exactly as it happens in the continuum
case~\cite{Brink:1976bc}.

The paper is organized as follows. Section~\ref{sec:qft} introduces
the main ingredients in the calculation of the quantum corrections to
the masses of the Kaluza--Klein zero modes, and derives the usual
formula for the mass shifts in a new effective field theory framework,
namely in the background field method with a Schwinger--time
regularization. The details of the string computation are described
in Section~\ref{sec:string}. Section~\ref{sec:latt} deals with the
details of the lattice computation, for the cases of bosonic and
fermionic contributions in the loops. The possibility of an accidental
extended supersymmetric is discussed at the end of
Section~\ref{sec:latt}. The main results of this work are summarized
in the conclusion together with some open questions that could be
addressed in future works.

\sect{Mass corrections in compactified field theories}
\label{sec:qft}

This section concentrates on the study of the mass renormalization in
the $SU(N)$ Yang--Mills gauge theory using the background field method
in a space--time with compactified dimensions. We shall see below that
even though our calculations are performed in a different setting,
they reproduce the results that have already appeared in the
literature. The correspondence between quantum field theory and string
computations is apparent when amplitudes are expressed in terms of
Schwinger parameters and an explicit mapping can be defined to relate
the string moduli and the Schwinger parameters,
see~\cite{Frizzo:2000ez} and references therein. We shall therefore
use the Schwinger parametrization in order to emphasize the connection
with the string theory approach. 
A similar approach has recently been developed in
Ref.~\cite{Ghilencea:2006qm,vonGersdorff:2008df}.

Clearly, before considering any explicit computation, the gauge
invariance of the theory should be used to constrain the form of the
2-gluon correlator. This is most easily done in a path integral
approach and by using BRS invariance, see for
instance~\cite{Collins}. In configuration space, this 2-point function
must satisfy the Ward identity
\begin{equation}\label{WI}
\frac{\partial}{\partial x^M} \frac{\partial}{\partial y^N}
\langle A^{a M}(x) A^{b N}(y) \rangle = -\ii \delta^{ab} 
\delta^{D} (x-y)~,
\end{equation}
where, in a $D$-dimensional theory, $M=0,\ldots, D-1$. If all dimensions
are uncompact this leads to the usual conclusion that the gluon
self-energy is transverse and no mass term can be generated. We
will see this feature arising explicitly in our 1-loop computation. In
a toroidal compactification the situation is different. Since we focus
on the case of vanishing Wilson lines, all fields are periodic around
the compact dimension and the associated momenta are discrete. Thus
for the Kaluza-Klein zero modes, Eq.~\eqref{WI} reduces to a constraint
involving only the gluon polarizations along the {\em uncompact}
directions, since these modes have a non-zero momentum only along
these directions. The mass correction of the other components (which
are scalars from the lower dimensional point of view) is not
constrained by any symmetry and can only be determined by 
performing an explicit computation. Let us notice that for the
{\em higher} Kaluza-Klein modes these Ward identities yield again
non-trivial constraints on the quantum mass corrections, see Section~3
of Ref.~\cite{Bauman:2007rt}, where this point is discussed in
detailed. 

Quantitative informations on the renormalization of zero-modes mass
can only be obtained by explicit calculations. We consider first the
case of a $D$-dimensional theory, without compact dimensions, in order
to set up our framework, and check indeed the symmetry constraints are
satisfied. 

Starting from the Feynman rules detailed in Ref.~\cite{Abbott:1980hw},
we compute the sum of one-loop diagrams contributing to the gauge
boson two-point function at zero external momentum\footnote{Note that
  we have used different metrics in different contexts. The field
  theory computation employs a ``mostly negative'' metric, the string
  theory computation a ``mostly positive'' metric, while the lattice
  computation is performed in Euclidean space--time. The reader should
  keep these conventions in mind in comparing results in this
  paper.}. In a $D$--dimensional theory, without compact dimensions,
there are four diagrams (as opposed to three in standard Yang--Mills
theories, as a result of an extra Feynman rule of two-ghost two-gluon
interaction), and their contributions are shown below:
\begin{align}
A_1=\parbox{40mm}{
\begin{center}
  \epsfig{figure=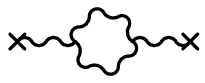}
\end{center}
}
=&2Dg_D^2 N\int\frac{d^Dk}{(2\pi)^D}\frac{k_\mu k_\nu}{k^4}~,\notag\\
A_2=\parbox{40mm}{
\begin{center}
  \epsfig{figure=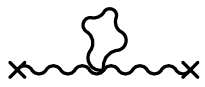}
\end{center}
}
=&-Dg_D^2 Ng_{\mu\nu}\int\frac{d^Dk}{(2\pi)^D}\frac{1}{k^2}~,\notag\\
A_3=\parbox{40mm}{
\begin{center}
      \epsfig{figure=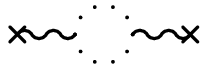}
\end{center}
}
=&-4g_D^2 N\int\frac{d^Dk}{(2\pi)^D}\frac{k_\mu k_\nu}{k^4}~,\notag\\
A_4=\parbox{40mm}{
\begin{center}
  \epsfig{figure=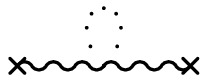}
\end{center}
}
=&2g_D^2 Ng_{\mu\nu}\int\frac{d^Dk}{(2\pi)^D}\frac{1}{k^2}~.
\end{align}
The sum of these amplitudes amounts to
\begin{equation}
  \label{eq:sum}
 A= (D-2)g_D^2N\int\frac{d^Dk}{(2\pi)^D}
  \left[\frac{2k_\mu k_\nu}{k^4}-\frac{g_{\mu\nu}}{k^2}\right]~.
\end{equation}
We suppress colour indices, which appear only in a delta function. 

This quantity is ultra-violet divergent and needs to be properly
regularized in order to evaluate the mass corrections. The gauge
coupling $g$ has mass dimension $[g]=\frac{4-D}{2}$, and so any
divergence contained in this amplitude which is to contribute to a
mass shift of the gauge boson must have mass dimension
$2-(4-D)=D-2$. In four dimensions this is a quadratic divergence.

In dimensional regularization, the divergence appears as a factor
$\Gamma\left(1-\frac{D}{2}\right)$, which has a first pole at $D=2$.
However, using the recursion relation for Gamma functions, this can be
transformed into a factor $\Gamma\left(2-\frac{D}{2}\right)$ (because
of the appearance of a $(D-2)$ factor before the integral) which has
its first pole at $D=4$ as expected for a logarithmic divergence in
four dimensions.

In this work a cutoff on the Schwinger time is used to regulate the
divergences. This is again in order to compare in a straightforward
way with perturbative string theory calculations, but also so that we
can extract the divergences as powers of a mass-scale $\Lambda$ which,
while we associate it with a momentum cut-off for the theory, does not
break the gauge invariance. This procedure involves the
exponentiation of the propagators in the momentum integrals using:
\begin{equation}
 \frac{1}{X^r}=\frac{1}{\Gamma(r)}\int_0^\infty dT\,T^{r-1}e^{-TX}~,
\end{equation}
where the variable $T$ is termed a Schwinger-time parameter. As an
example, this procedure yields for a tadpole diagram:
\begin{equation}
  \int\frac{d^Dk}{(2\pi)^D}\frac{1}{k^2}=\int_0^\infty
  dT\,\int\frac{d^Dk}{(2\pi)^D}e^{-Tk^2}
  =\frac{1}{(4\pi)^\frac{D}{2}}\int_0^\infty dT\, T^{-\frac{D}{2}}~,
\end{equation}
which is a divergent integral. The divergences arise from the
$T\rightarrow0$ region of the integral, where there is no exponential
damping of the contribution from large momenta in the above
expression, and so we can regulate by imposing a lower bound $T_0$ on
the integration variable $T$. Doing this we see that the divergence
appears in the result as a factor $T_0^{1-\frac{D}{2}}$ and so, as we
expect, this divergence is of mass dimension $D-2$. In order to
associate the lower bound on the Schwinger-time with a momentum
cut-off $\Lambda$, we write $T_0=\frac{1}{\Lambda^2}$. Thus the two
integrals contained in Eq.~(\ref{eq:sum}) amount to
\begin{eqnarray}
  \int\frac{d^Dk}{(2\pi)^D}\frac{1}{k^2}=
  \frac{-\ii}{(4\pi)^\frac{D}{2}}\frac{\Lambda^{D-2}}{\frac{D}{2}-1}
  ~~,&~~~& 
  \int\frac{d^Dk}{(2\pi)^D}\frac{k_\mu k_\nu}{k^4}=\frac{-\ii g_{\mu\nu}}
  {2(4\pi)^\frac{D}{2}}\frac{\Lambda^{D-2}}{\frac{D}{2}-1}~.
\end{eqnarray}
Inserting this result into Eq.~(\ref{eq:sum}) we see that the two
terms cancel and the expression vanishes as required by gauge
invariance.

Let us now examine the effects of compactification on this
cancellation. We restrict ourselves to the case where we compactify
one of the $D$ dimensions, leaving an effective theory in $d=D-1$
dimensions. The resulting effective theory consists of a $d$ component
gauge boson, and a scalar field in the adjoint representation arising
from the extra-dimensional component of the original $D$ component
gauge field. The momentum of the fields in the finite compactified
dimension produces a tree--level mass for an infinite tower of fields
called Kaluza-Klein (KK) modes. The gauge coupling is rescaled by
$g_d^2=\frac{g_D^2}{2\pi R}$.

The zero mode gauge boson does not receives any 1--loop mass
renormalization after compactification since the computation of the
2--point function is basically the one discussed above. The adjoint
scalar however, does, as expected due to the breaking of the original
gauge invariance. We will illustrate this here, and confirm agreement
with the result obtained in
Ref.~\cite{Hatanaka:1998yp,vonGersdorff:2002as,Cheng:2002iz}. Note
that in Ref.~\cite{Cheng:2002iz} the relevant two-point functions are
computed not at zero external momentum $p$, but in the approximation
$p^2=r^2$ where $r=p_5=\frac{n}{R}$ is the KK mass of the external
particle. As a result of the Poisson Resummation used to compute the
sum over KK modes, inverse powers of the KK mode of the external
particle are generated, which can yield extra contributions to the
final result which would be missed in the $p=0$ limit. This only
affects the result for $r\neq0$ external modes however, and therefore
we can work at $p=0$ for our purposes.

By keeping a generic non-vanishing external momentum $p\neq0$ for the
zero modes, it would be possible to compute higher-derivative terms in
the low-energy effective action which can be relevant in
phenomenological applications~\cite{Ghilencea:2004sq}. However, in
this paper, we focus on the mass correction terms which represent the
most relevant contributions in the infrared, and which are most easily
computed both in string theory and in the lattice field theory
approach.

In computing the contributions to the scalar two-point function at
zero momentum, there are two integrals, summed over Kaluza-Klein
modes, which arise. These are
\begin{eqnarray}
\label{eq:integrals}
  \mathcal{I}_1 & = & \sum_l\int\frac{d^dk}{(2\pi)^d}\frac{1}{k^2-l^2}
  = -\frac{\ii}{(4\pi)^{d/2}} \sum_l
  \int_0^\infty \frac{dT}{T^{\frac{d}{2}}}\; \ex{-l^2 T}~, 
  \\ \nonumber
  \mathcal{I}_2 & = &
  \sum_l\int\frac{d^dk}{(2\pi)^d}\frac{l^2}{(k^2-l^2)^2} 
  = \frac{\ii}{(4\pi)^\frac{d}{2}} \sum_ll^2
  \int_0^\infty \frac{dT}{T^{\frac{d}{2}-1}} \; \ex{-l^2 T}~, 
\end{eqnarray}
with $l=\frac{m}{R}$ where $m$ is an integer denotes the mass of a KK
mode, and the sum over $l$ is a sum over the integers $m$.

There are seven diagrams contributing to the two-point function for
the adjoint scalar field; they are shown in Tab.~\ref{fig:7diag} with
their contributions in terms of $\mathcal{I}_1$ and $\mathcal{I}_2$.
\begin{table}[H]
\centering
\begin{tabular}{|c|c|c|}\hline
Diagram & $g_d^2N\mathcal{I}_1$&$g_d^2N\mathcal{I}_2$\\\hline
\parbox{25mm}{
\begin{center}
  \epsfig{figure=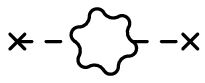}
\end{center}
}
&0&$2d$\\\hline
\parbox{25mm}{
\begin{center}
  \epsfig{figure=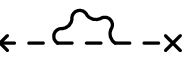}
\end{center}
}
&0\label{zero}&0\\\hline
\parbox{25mm}{
\begin{center}
  \epsfig{figure=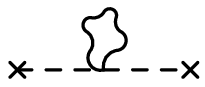}
\end{center}
}
&$d$&0\\\hline
\parbox{25mm}{
\begin{center}
  \epsfig{figure=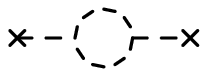}
\end{center}
}
&0&$2$\\\hline
\parbox{25mm}{
\begin{center}
  \epsfig{figure=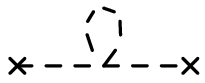}
\end{center}
}
&$1$&0\\\hline
\parbox{25mm}{
\begin{center}
  \epsfig{figure=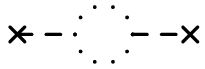}
\end{center}
}
&0&$-4$\\\hline
\parbox{25mm}{
\begin{center}
  \epsfig{figure=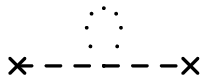}
\end{center}
}
&$-2$&0\\\hline
\end{tabular}
  \caption{Diagrams yielding the quantum corrections to the adjoint scalar mass.}
  \label{fig:7diag}
\end{table}
The second diagram in Tab.~\ref{fig:7diag} vanishes only at zero
momentum and in the Feynman type gauge\footnote{i.e. $\alpha$ the parameter of the background gauge fixing term is $\alpha=1$, see \cite{Abbott:1980hw}}. The
sum of all the diagrams produces
\begin{equation}
 A=(d-1)g_{d}^2N[\mathcal{I}_1+2\mathcal{I}_2] \, .
\label{eq:compact}
\end{equation}

We evaluate the integrals from (\ref{eq:integrals}) following a
similar procedure to the non-compact case. After exponentiating
propagators and performing the Gaussian momentum integral, we find
quantities such as $\displaystyle\sum_{l}e^{-Tl^2}$ where the sum is over the KK
mode of the loop particle. We evaluate such infinite sums through a
Poisson resummation, transforming the sum over KK modes into a sum
over the winding number of the path of the loop particle around the
compact extra dimension.
\begin{equation}
 \sum_{l=\frac{m}{R}}e^{-Tl^2}=\frac{(2\pi R)}{\sqrt{4\pi T}}\sum_{n}e^{-\frac{\pi^2R^2 n^2}{T}}
\end{equation}
The $n=0$ term corresponds to the noncompact case, and so produces a
$\Lambda^{D-2}$ divergence. In all other terms we make the change of
variables $t=\frac{\pi^2 R^2 n^2}{T}$ which then results in the
integration over the Schwinger parameter producing Gamma functions
with arguments away from the singularities. Omitting further details,
we obtain
\begin{align}
  \mathcal{I}_1=&-\frac{\ii}{2}\frac{(2\pi R)^{2-d}}{\pi^{\frac{d+1}{2}}}
  \zeta(d-1)\Gamma\left(\frac{d-1}{2}\right)-\frac{i}{(4\pi)^\frac{d}{2}}\frac{2\sqrt{\pi}R}{d-1}\Lambda^{d-1}~,\notag\\
  \mathcal{I}_2=&\frac{\ii}{2}\frac{(2\pi R)^{2-d}}{\pi^{\frac{d+1}{2}}}\zeta(d-1)
  \left[\frac{1}{2}\Gamma\left(\frac{d-1}{2}\right)-
    \Gamma\left(\frac{d+1}{2}\right)\right]+\frac{1}{2}\frac{i}{(4\pi)^\frac{d}{2}}\frac{2\sqrt{\pi}R}{d-1}\Lambda^{d-1}~ .
\end{align}
In Tab.~\ref{fig:7diagexp} we show the final contribution of each
diagram. We quote the coefficient of
$\frac{\ii}{(4\pi)^\frac{d}{2}}\frac{2}{\sqrt{\pi}}(\pi
R)^{2-d}\zeta(d-1)\Gamma\left(\frac{d-1}{2}\right)$, and also
$\frac{\ii}{(4\pi)^\frac{d}{2}}\frac{2\sqrt{\pi}R}{d-1}\Lambda^{d-1}$
that results for each diagram

\begin{table}[H]
  \centering
\begin{tabular}{|c|c|c|}\hline
Diagram & $\sim R^{2-d}$&$\sim \Lambda^{d-1}$\\\hline
\parbox{25mm}{
\begin{center}
  \epsfig{figure=drft9.eps}
\end{center}
}
&$d(2-d)$&$d$\\\hline
\parbox{25mm}{
\begin{center}
  \epsfig{figure=drft10.eps}
\end{center}
}
&$-d$&$-d$\\\hline
\parbox{25mm}{
\begin{center}
  \epsfig{figure=drft11.eps}
\end{center}
}
&$2-d$&1\\\hline
\parbox{25mm}{
\begin{center}
  \epsfig{figure=drft12.eps}
\end{center}
}
&-1&-1\\\hline
\parbox{25mm}{
\begin{center}
  \epsfig{figure=drft13.eps}
\end{center}
}
&$-2(2-d)$&-2\\\hline
\parbox{25mm}{
\begin{center}
  \epsfig{figure=drft14.eps}
\end{center}
}
&2&2\\\hline
\end{tabular}  
  \caption{Non-zero contributions from the diagrams in
    Fig.~\ref{fig:7diag}}
  \label{fig:7diagexp}
\end{table}

It is easily seen that the divergent contributions cancel each other,
as expected by the higher dimensional gauge invariance. The total
contribution then becomes
\begin{align}
A=&(d-1)g_{d}^2N\sum_{l}\int\frac{d^{d}k}{(2\pi)^{d}}
  \left[\frac{2l^2}{(k^2-l^2)^2}+\frac{1}{k^2-l^2}\right]\notag\\
  =&-(d-1)g_{d}^2N\frac{\ii(2\pi
    R)^{2-d}}{\pi^{\frac{d+1}{2}}}\zeta(d-1)
  \Gamma\left(\frac{d+1}{2}\right)\, .
\end{align}
This results in an additive mass renormalization of the adjoint scalar
field of
\begin{equation} 
 \label{eq:deltams}
 \delta
 m^2 =\frac{g_{d}^2N}{\pi^{\frac{d+1}{2}}}\frac{(D-2)}{(2\pi R)^{d-2}}\zeta(d-1)
 \Gamma\left(\frac{d+1}{2}\right)~ .
\end{equation}
For $D=5$ and $d=4$, this gives
\begin{equation}
  \delta m^2=\frac{9g_4^2N}{16\pi^4 R^2}\zeta(3)\, 
\end{equation}

\sect{Mass corrections in the open bosonic string theory}
\label{sec:string}

The naive vacuum of the open bosonic string theory is unstable, as it
is signalled by the presence of a tachyon excitation with mass square
$M^2_{\rm t}=-1/\alpha'$. However, it is still useful to study
formally the perturbation theory around this unstable point, since in
this way it is possible to understand, in a simple setup, many
properties of the string amplitudes of fully consistent theories. In
practice, one can compute the loop amplitudes by using standard string
techniques~\cite{Green:1987mn} and discard by hand the tachyon
contributions before considering the loop integrals. This approach has
been already used successfully in the past in the study of the low
energy limit of 1--loop string amplitudes, see for
instance~\cite{Bern:1992ad,Frizzo:2000ez}. Moreover the analysis of
the radiative corrections to the mass of the string states was
initiated in the context of bosonic
theory~\cite{Weinberg:1985je}. Most of the early studies of these
radiative corrections were done in the context of closed string
theory~\cite{Seiberg:1986ea,Sen:1987bd,Minahan:1989cb}. More
recently~\cite{Antoniadis:2000tq,Kitazawa:2008tb}, the same problem
has been analyzed in an open string context by computing the effective
action for two stacks of D-branes. In this section we will consider
this (bosonic) D-brane setup, but we will follow the original approach
of~\cite{Weinberg:1985je} and compute the 2-point function for open
strings on the annulus. Even if we focus on the string states
corresponding to the internal components of the gauge field, this
approach can be used also when the vertex operators contain twist
fields, a situation where the technique used
in Refs.~\cite{Antoniadis:2000tq,Kitazawa:2008tb} cannot be applied.

Let us consider a stack of $N$ space-filling D-branes in bosonic
string theory and we take the spacetime to be the product of the
$d$-dimensional Minkowski space and $s$ circles of radius $R$ (in
principle, bosonic string is critical only if $d+s\equiv D =26$,
however this constraint will play no role in most of our
computations). We will focus on the massless open string states
supported by these D-branes. The (onshell) 2-point amplitude with
massless states requires to take the external particles at zero
momentum, which is sufficient for computing the 1--loop mass
corrections we are interested in. The vertex operator describing these
states is simply\footnote{We use the same conventions
  of Ref.~\cite{DiVecchia:1996uq}. Eq.~\eqref{VO} basically states that
  open string endpoints are minimally coupled to the gauge field. This
  fixes also the overall normalization of the vertex
  operator. Alternatively the normalization can be determined by using
  unitarity and by matching the low energy behavior of the tree-level
  3-point function against the Yang--Mills 3-gluon vertex.}
\begin{equation}
  \label{VO}
V^a = \ii {g_D} T^a \partial X^I~, 
\end{equation}
where $T^a$ is a $SU(N)$ generator\footnote{At the full string level
the gauge group is $U(N)$, however all amplitudes with external $U(1)$
massless states vanish.}, $g_D$ is the $D$-dimensional Yang-Mills
coupling and $I\equiv \mu=0,\ldots,d-1$ for the vector boson, while
$I\equiv i=1,\ldots,s$ for the scalars arising from the Kaluza-Klein
reduction of the higher dimensional gauge field.

The radiative correction to the tree-level mass square ($\delta
m^2_I$)  is obtained from the planar 2-point amplitude $A^I$: 
\begin{equation}
  \label{massStr}
  A^I = -{g_D^2 N} \Tr(T^a T^b) \int \langle 
\partial X^I(1) \partial X^I(y) 
\rangle~ d\mu\, ,
\end{equation}
where in our conventions $\Tr(T^a T^b)={\delta^{ab}}/2$ and of course
the index $I$ is not summed, but takes one of the values listed
above. In our case the correlator $\langle\ldots \rangle$ is taken
over the annulus topology, $d\mu$ is the 1--loop integration
measure~\eqref{dmu}. Let us analyze these ingredients in some detail.

We will parametrize the annulus as the upper half complex plane (minus
the point $z=0$) modded out by the equivalence relation $z \to k z$,
where $k$ is a real number $k\in (0,1)$. Each value of $k$ correspond
to a different shape for the annulus and, in the
amplitude~\eqref{massStr}, we need to integrate over all
possibilities. The two borders of the annulus are the segments on the
real axis $y\in[k,1]$ and $y\in[-1,-k]$. We are free to choose the
position of the first vertex operator and the second vertex operator
has to stay on the same border, $y\in [k,1]$ in our case. The
correlator $\langle \partial X^I(1) \partial X^I(y) \rangle$ can be
split in the contribution of the vibration modes of the string and the
one of the center of mass and rigid motion (zero modes). By following
the derivation in Chapter~8 of~\cite{Green:1987mn}, one can compute
these correlators. The non zero--mode part is expressed in terms of the
Green function satisfying Dirichlet boundary conditions $G_D$
\begin{equation}
  \label{nzm}
  \langle \partial X^I(y_1) \partial X^J(y_2) \rangle_{\rm nzm}
  = -2\alpha' \eta^{IJ} \partial_{y_1} \partial_{y_2} G_D(y_1,y_2)~
\end{equation}
with
\begin{equation}
  \label{DirG}
   G_D(y_1,y_2) =  \ln \left[(y_1 -
    y_2) \prod_{n=1}^\infty \frac{(1-k^n y_1/y_2) (1-k^n
  y_2/y_1)}{(1-k^n)^2}\right] - \frac 12 \ln y_1 - \frac 12 \ln y_2 ~, 
\end{equation}
where the last two terms have been added so that $G_D$ has simpler
periodicity properties, but obviously they do not contribute
to~\eqref{nzm}. In the computation of the zero mode part we use the
expansion $\partial X(y) = -\ii (2\alpha')\hat{p}/y + \ldots$, where
the dots stand for the non zero-mode we have already taken into
consideration. Thus we get
\begin{equation}
  \label{corr1}
  \langle \partial X^I(y_1) \partial X^J(y_2) \rangle_{zm} =
  \frac{Vol}{(2\pi R)^s} \sum_{n^i} \int   \frac{d^dp}{(2\pi)^d}
  \left[-(2\alpha')^2\frac{p^I}{y_1} \frac{p^J}{y_2}+\langle 
    \ldots \rangle_{\rm nzm} \right]  \ex{\alpha'
  (\sum p_\mu^2+\sum_i \frac{n_i^2}{R^2}) \ln k}~,
\end{equation}
where the volume is $Vol=(2\pi)^D \delta^d(0) R^s$. If we consider
standard gauge bosons as external states, the index $I$ lies in the
non-compact space. After integrating over $p$, we can see that the
zero-mode contribution combines with the non zero-mode one and
transforms the Dirichlet Green function into the Neumann one $G_N$:
\begin{equation}
  \label{NeuG}
  G_N(y_1,y_2) = G_D(y_1,y_2) + \frac{(\ln y_1 - \ln y_2)^2}{2 \ln
    k}~.
\end{equation}

Let us analyze the gauge boson mass corrections first and show that we
get a vanishing mass correction as required by gauge invariance. The
1--loop measure is
\begin{equation}
  \label{dmu}
  \frac{dk}{k^2} dy~ [\mu(k)] = \frac{dk}{k^2} dy
  \left[ \prod_{n=1}^\infty (1-k^n)^{2-D}
  \right] \left(-\frac{\pi}{\alpha'\ln k}\right)^{d/2}~,
\end{equation}
where the last factor follows from the Gaussian integration
in~\eqref{corr1} and the product over $n$ is the contribution of the
string vibration modes. Then, from~\eqref{massStr} we read
\begin{equation}
  \label{mmu}
  \delta m^2_\mu = -{\alpha'} \frac{g_d^2 N}{(2\pi)^d} 
  \int_0^1 dk ~[\mu(k)] \int_k^1 dy 
  \left[\theta\left(0\Big|\frac{-\ii \alpha'\ln k}{\pi R^2}
  \right)\right]^s 
  \partial_{y_1} \partial_y G_N(y_1,y)\Big|_{y_1=1}~,
\end{equation}
where $g_d$ is the d-dimensional Yang--Mills coupling
$g_d^2=g_D^2/(2\pi R)^s$ and
\begin{equation}
  \label{thetac}
\theta(\nu|\tau)= \sum_n \exp[\pi\ii n^2\tau+2\pi \ii n \nu]~.
\end{equation}
Apparently $\delta m^2_\mu$ is trivially zero, since the integrand is
a total derivative. However, as discussed in Ref.~\cite{Minahan:1989cb},
one has to keep in mind two points: first the integrand is
quadratically divergent as $y\to 1$ or $y\to k$ so it has to be
regularized, then after regularization~\eqref{mmu} is zero only if the
integral over $y$ is single valued on the boundary of the annulus
(i.e. periodic when $y \to k y$). By using the explicit expression for
the Green functions~\eqref{DirG} and~\eqref{NeuG} one can check the
following properties
\begin{equation}
\label{Npe}
\partial_{y_1} {G}_{N}(y_1,k y_2) = \partial_{y_1} {G}_{N}(y_1,y_2)
=  \partial_{y_1} {G}_{N}(y_2,y_1) ~~,~~~~ 
{G}_{N}(y_2^{-1}, y_1^{-1}) = {G}_{N}(y_1,y_2)~.
\end{equation}
Then we regularize the integral~\eqref{mmu} simply by cutting away the
dangerous region around $y=1\sim k$ and, by using~\eqref{Npe}, we get
\begin{equation}
  \label{regy}
 \int_k^1 dy ~ \partial_y \partial {G}_{N}(1,y) \to
\int\limits_{k/(1-\epsilon)}^{1-\epsilon} dy ~ \partial_y \partial
{G}_{N}(1,y) = 2 \partial {G}_{N}(1,1-\epsilon) \sim
\frac{2}{\epsilon} + {\cal O}(\epsilon)~.
\end{equation}
The divergent contribution is due the exchange of an off-shell
zero-momentum tachyon. It can be renormalized away by redefining
the 2-dimensional cosmological constant, that is by adding to the
world-sheet sigma model a coupling $C \int \sqrt{h}$, where $h$ is the
metric on the world-sheet and $C$ is an appropriate constant. As
usual~\cite{Green:1987sp}, we will discard this divergent contribution
without leaving any additional finite part. After this regularization
Eq.~\eqref{regy} vanishes and no radiative mass correction for the
gauge boson is generated at 1--loop.

The situation is very different if we consider the scalars arising
from the Kaluza-Klein compactification $I=i=1,\ldots,s$. Let us first
focus on the case $R\to 0$, where the analysis simplifies (by applying
a T-duality this limit is equivalent to a lower dimensional D-brane in
the uncompact space). In this limit the sum in Eq.~\eqref{corr1}
vanishes and thus we have:
\begin{equation}
  \label{mi0}
  \delta m^2_i(R\to 0) = -{\alpha'} \frac{g_d^2 N}{(2\pi)^d} 
  \int_0^1 dk ~[\mu(k)] \int_k^1 dy 
  \partial_{y_1} \partial_y G_D(y_1,y)\Big|_{y_1=1}~.
\end{equation}
By using Eq.~\eqref{NeuG} we can see that the integral over $y$ now
yields also a finite term
\begin{equation}
  \label{regyD}
\int\limits_{k/(1-\epsilon)}^{1-\epsilon} dy ~ \partial_y \partial
{G}_{D}(1,y) = \int\limits_{k/(1-\epsilon)}^{1-\epsilon} dy ~
\partial_y \left[\partial{G}_{N}(1,1-\epsilon) +\frac{\ln y}{\ln
k}\right] \sim \frac{2}{\epsilon} - 1 + {\cal O}(\epsilon)~.
\end{equation}
Thus, by implementing the same subtraction used in the gluon case,
we are left with a non-zero contribution to the scalar mass
\begin{equation}
  \label{mi}
  \delta m^2_i(R\to 0) = {\alpha'} \frac{g_d^2 N}{(2\pi)^d}
  \int_0^1 \frac{dk}{k^2}
  \left[ \prod_{n=1}^\infty (1-k^n)^{2-D} \right]
\left(\frac{-\pi}{\alpha'\ln k}\right)^{d/2}~.
\end{equation}
This result is still divergent as $k\to 0$ (and $k\to 1$), but these
are physical poles that correspond to the propagation of the open (and
closed) string tachyon. In a tachyon-free string theory these poles
will be automatically absent, in the present case we will subtract
them by hand.  

Let us now consider the case of a compactification with finite radius.
\begin{eqnarray}
  \label{miR0}
  \delta m^2_i & =& -{\alpha'} \frac{g_d^2 N}{(2\pi)^d}
 \int_0^1 \frac{dk}{k^2} \prod_{n=1}^\infty (1-k^n)^{2-D} 
 \left(-\frac{\pi}{\alpha'\ln k}\right)^{d/2} 
  \left[\theta\left(0\Big|\frac{-\ii \alpha'\ln k}{\pi R^2}
  \right)\right]^{s-1}  
\\ \nonumber & \times & \int_k^1 dy 
\left[\partial \partial_y {G_D}(1,y)~
\theta\left(0\Big| -\frac{\ii\alpha'}{\pi R^2} \ln k\right)
-\frac{2 \alpha'}{4\pi^2 R^2} \frac{1}{y} ~ \theta''\left(0\Big|
 -\frac{\ii\alpha'}{\pi R^2} \ln  k\right) \right]~,
\end{eqnarray}
where $\theta'(0|\tau)=\partial_\nu \theta(\nu|\tau)|_{\nu=0}$. By
using the regularization~\eqref{regyD}, the integral over $y$ can be
performed explicitly. Then one can see that the first term is the
stringy generalization of the field theory term proportional to ${\cal
  I}_1$, while the second one generalizes the contribution $2 {\cal
  I}_2$ in Eq.~\eqref{eq:compact}. Both integrands are now dressed
with the Dedekind function~\eqref{dedekind} $\eta$~function which
takes into account the contribution of the stringy modes. In order to
compute the mass shift it is convenient to invert the modular
parameter in~\eqref{miR0} so that the two terms combine in a single
contribution. Under this transformation , the $\theta$ function
transforms as follow
\begin{equation}\label{pr}
\theta(\nu|\tau) = \frac{1}{\sqrt{-\ii \tau}} \ex{-\pi\ii\nu^2/\tau} 
\theta\left(\frac{\nu}{\tau}\Big|-\frac{1}{\tau}\right)~,
\end{equation}
which implies
\begin{equation}\label{prd2}
\theta''(0|\tau) = \frac{1}{\sqrt{-\ii \tau}} \frac{1}{\tau^2}
\theta''\left(0\Big|-\frac{1}{\tau}\right) - \frac{1}{\sqrt{-\ii
\tau}} \frac{2\pi\ii}{\tau} \theta\left(0\Big|-\frac{1}{\tau}\right)~.
\end{equation}
By using~\eqref{pr} and~\eqref{prd2} in~\eqref{miR0}, we can combine
the terms proportional to $\theta$ and reconstruct again the Neumman
Green function~\eqref{NeuG}. As we have seen above, in this case the
(properly regularized) integral over $y$ vanishes. Thus 
\begin{equation}
  \label{miR}
  \delta m^2_i  =
  \frac{-g_d^2 N}{(2\alpha')^{\frac{d}{2}-2}}
  \frac{1}{(2\pi)^{d+1}} \frac{1}{2 R^2}
 \int_0^1 \frac{dk}{k^2} \prod_{n=1}^\infty (1-k^n)^{2-D} 
 \left(\frac{-2\pi}{\ln k}\right)^{\frac{d}{2}-1} 
  \left[\theta\left(0\Big|\ii T_R \right)\right]^{s-1}
  \frac{\theta''(0\Big|\frac{\ii}{T_R})}{T_R^{\frac{5}{2}}}~,
\end{equation}
where $T_R = -(\alpha' \ln k)/(\pi R^2)$. If we work with a critical
theory ($D=26$) and consider the case of a single compact dimension
($s=1$), we recover Eq.~(81) of~\cite{Kitazawa:2008tb}. In order to
match the results, one need to perform a modular transformation and use 
\begin{equation}
  \label{dedekind}
  k^{1/24} \prod_{n=1}^\infty (1-k^n) =\eta\left(\frac{\ln
      k}{2\pi\ii}\right) = \left(\frac{-2\pi}{\ln k}\right)^{1/2}
  \eta\left(-\frac{2\pi\ii}{\ln k}\right)~. 
\end{equation}
Then in this case we can write~\eqref{miR} in the closed string
channel $t_c=-1/\ln k$ and we obtain\footnote{Contrary to what is
  claimed in~\cite{Kitazawa:2008tb}, Eq.~\eqref{kitazawa} does not
  vanish in the limit $R\to 0$. In this limit the sum over $w$ becomes
  an integral and one recovers~\eqref{mi}.}
\begin{equation}
  \label{kitazawa}
  \frac 12 \frac{g_{26}^2}{(2\alpha')^{13}}
  \frac{(2\pi R)^2}{(2\pi)^{25}} 
 \int_0^\infty \!dt_c~ \eta^{-24}(2\pi\ii t_c)
 \sum_{w=-\infty}^\infty w^2 \ex{-w^2 t_c \pi^2 R^2/\alpha' }~.
\end{equation}

Let us go back to Eq.~\eqref{miR} and study the compactification on a
circle ($s=1$) for a generic dimension $d$. If we discard by hand the
tachyon poles, the leading contribution comes from the region $T_R >
1$ (i.e. $|\ln k| > \pi R^2/\alpha'$) where the $\theta''$ is not
suppressed. In the regime where the string scale is much higher than
the compactification scale $R^2 \gg \alpha'$, this implies also $|\ln
k| \gg 1$. In this limit the string amplitudes reduce to the field
theory result, see~\cite{Frizzo:2000ez} and references therein.
Thus we expect that, when $R^2\gg \alpha'$, the string result
automatically reduces to the field theory one~\eqref{eq:deltams}. Let
us check that this is indeed the case. Since $k$ is small we can
expand the product over $n$ in~\eqref{miR} and keep only the second
term that cancels the tachyonic pole. Then we have
\begin{equation}
  \label{mi2}
  \delta m^2_i  \sim
  \frac{g_d^2 N}{(2\alpha')^{\frac{d}{2}-2}}
  \frac{1}{(2\pi)^{d-2}} \frac{D-2}{2 R^2}
  \left(\frac{2\alpha'}{R^2}\right)^{\frac d2 -2}
  \int_0^\infty dT_R T_R^{-\frac{3}{2}-\frac{d}{2}} 
  \sum_{w=-\infty}^{\infty} w^2 \ex{-\pi w^2/T_R}~.
\end{equation}
By means of a change of variable the integral reduces to the Euler
formula of the Gamma function and the sum to the definition of the
Riemann zeta function
\begin{equation}
  \label{strft}
    \delta m^2_i  \sim
  \frac{g_d^2 N}{\pi^{\frac{d+1}{2}}}
  \frac{D-2}{(2\pi R)^{d-2}} 
  \Gamma\left(\frac{d+1}{2}\right)
  \zeta\left(d-1 \right)\, .
\end{equation}

Let us close this section by noting that 
the
mechanism we have just discussed actually holds at any order in
perturbation theory. The explicit expressions for the the Green
functions with Dirichlet or Neumann boundary condition become more
involved on a world-sheet with an arbitrary number of holes or
handles. However everything can be written in terms of classical
functions defined on the appropriate Riemann surface, such as the
Abelian differentials and the prime form (see for
instance~\cite{DiVecchia:1988cy}). The main ingredient used in the
string computation is the periodicity of the integrand when the
relative position of the two vertex operators changes. It is possible
to generalize step by step the procedure described in this section and
check that even with the higher loop Green functions the integral over
the relative position of the punctures yields the same results as
in~\eqref{regy} and~\eqref{regyD}. Thus the the vector states are
protected against a mass renormalization because in the relevant
string 2-point function the Neumann Green function appears. On the
contrary, the internal polarizations of the gauge field are not
protected and the higher loop contributions to the mass shift are
given by a generalization of~\eqref{miR} which involves Riemann's
$\theta$-functions instead of the Jacobi's ones. Still we expect that
the same mechanism described above is at work: when $R^2\gg \alpha'$
the elements of the period matrix, which generalize the 1--loop
parameter $\ln k$, must be large otherwise the result is
suppressed. In this limit, we expect\footnote{When one wants to focus
  on the contributions from the massless states in the loops, as we
  have done in~\eqref{mi2}, it is rather difficult to explicitly check
  this point even at two loops.} that the string answer reduces to the
field theory one and all factors of $\alpha'$ cancel. 
At first sight this seems to be in agreement with results obtained at
two--loop in quantum field
theory~\cite{Hosotani:2007kn,Delgado:2001xr}. A more careful
investigation is needed in order to clarify this issue.

\sect{Mass corrections on the lattice.}
\label{sec:latt}

In this section we consider $(d+1)$-dimensional gauge theories
regularized on an asymmetric lattice. In particular, we consider one
dimension to be much smaller than the remaining ones so as to recover in
the continuum limit a theory compactified on a circle. We use $N_s$ to
indicate the number of points in the compact dimension, then its
radius $R$ is $2\pi R= N_s a$, where $a$ is the lattice spacing.  By
using standard lattice perturbation theory, we compute (again) the
1--loop radiative corrections to the mass of the gluon and the scalar
states. A similar computation for standard four dimensional theories
 was performed in Ref.~\cite{Kawai:1980ja} in order to check that there is
no mass renormalization for the vector bosons, as required by gauge
invariance. We want to see how this result changes when the finite
size effects of the $S^1$ compactification are taken into account. In
this case, gauge invariance does not protect the mass of the gauge
boson polarized along the $S^1$. The compactification from four to
three dimensions has been analyzed in detail
in Ref.~\cite{Kaste:1997ks}. As already discussed in the introduction, this
case is relevant for studying the thermal behavior of Yang--Mills: the
non-zero mass corrections to the time component of the vector boson
are interpreted as a screening effect for the (electric) components of
the force.

\subsection{Bosonic contribution}

\begin{figure}[ht]
  \centering
  \begin{tabular}[h]{lc}
(a) gluon sunset: &
    \parbox{.8truein}{ 
      \epsfig{figure=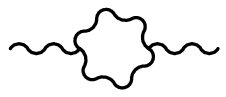}
      } \\
(b) gluon tadpole: &
    \parbox{.8truein}{ 
      \epsfig{figure=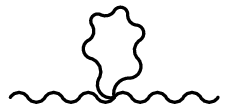}
      } \\
(c) measure: &
    \parbox{.8truein}{ 
      \epsfig{figure=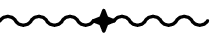}
      } \\
(d) ghost sunset: &
    \parbox{.8truein}{ 
      \epsfig{figure=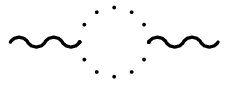}
      } \\
(e) ghost tadpole: &
    \parbox{.8truein}{ 
      \epsfig{figure=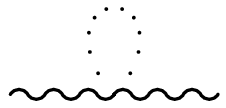}
      } \\
\end{tabular}
\caption{Contributions to the gluon two--point function in the pure
  gauge theory.  }
  \label{fig:latdiag}
\end{figure}

We start by focusing on pure Yang--Mills theory. Technically we mix the
approaches of~\cite{Kawai:1980ja} and~\cite{Kaste:1997ks}. We compute
the five 1--loop Feynman diagrams contributing to the 2-point
function, see Fig.~\ref{fig:latdiag}, when one of the $D$ dimensions
of the lattice is compact, and focus on the component of the 2-point
function in this direction in order to examine the mass correction to
the generated adjoint scalar. Since we are interested only in
extracting the mass correction we can put the external momenta to
zero, which simplify drastically the computation with the 4-particle
vertices. Then we combine these contributions together by using a
discrete version of the partial integration introduced
in Ref.~\cite{Kawai:1980ja}. Let us see how this works in details.

By using the Feynman rules listed in Refs.~\cite{Kawai:1980ja,Kaste:1997ks},
it is straightforward to construct the contribution of the diagram in
Fig.~\ref{fig:latdiag}.a:
\begin{equation}
\label{sunsetglu}
A_a = \frac{g^2_{d+1}N}{a^{d-1}} \frac{1}{N_s} \sum_{n=0}^{N_s-1}
\int_{-\pi}^\pi \frac{d^dq}{(2\pi)^d} \left\{ \frac{\cos^2 \frac{\pi
      n}{N_s}}{D_n} + \left[\frac{d-2}{2}+ \frac 14 \right]
  \frac{\sin^2 \frac{2\pi n}{N_s}}{D_n^2}\right\} ~,
\end{equation}
where $g_{d+1}$ is the Yang--Mills coupling of the higher dimensional
theory and $1/D_n$ is the bosonic propagator
\begin{equation}
\label{pro}
D_n = 4 \sin^2\frac{\pi n}{N_s} + 4 \sum_{i=1}^d \sin^2\frac{q_i}{2}~.
\end{equation}
The factor of $1/4$ in the square parenthesis of Eq.~\eqref{sunsetglu}
cancels against the ghost loop depicted in Fig.~\ref{fig:latdiag}.d:
\begin{equation}
\label{sunsetgho}
A_d = \frac{g^2_{d+1}N}{a^2} \frac{1}{N_s} \sum_{n=0}^{N_s-1} 
\int_{-\pi}^\pi \frac{d^dq}{(2\pi)^d} \left(-\frac 14
\frac{\sin^2 \frac{2\pi n}{N_s}}{D_n^2}\right)~.
\end{equation}
Let us now consider the tadpole contributions
(Fig.~\ref{fig:latdiag}.b and~\ref{fig:latdiag}.e):
\begin{equation}
\label{tadglu}
A_b = \frac{g^2_{d+1}N}{a^{d-1}} \frac{1}{N_s} \left[ \frac{1}{12}
-\sum_{n=0}^{N_s-1}  \int_{-\pi}^\pi \frac{d^dq}{(2\pi)^d} 
\frac{d \cos\frac{2\pi n}{N_s}-\cos^2\frac{\pi n}{N_s}+\frac{4}{3}
\sin^2\frac{\pi n}{N_s}}{D_n}\right]~
\end{equation}
for the gluon loop, and
\begin{equation}
\label{tadgho}
A_e = -\frac{2}{3} \frac{g^2_{d+1}N}{a^{d-1}} \frac{1}{N_s} \sum_{n=0}^{N_s-1}  
\int_{-\pi}^\pi \frac{d^dq}{(2\pi)^d}
\frac{\sin^2\frac{\pi n}{N_s}}{D_n}
\end{equation}
for the ghost loop. The factor of $1/12$ in~\eqref{tadglu} cancels
against the the diagram in Fig.~\ref{fig:latdiag}.c, which arises in
the lattice regularization from the integration measure. Thus, by
combining all these contributions, we obtain a simple expression for
the gluon for the complete amplitude:
\begin{equation}
\label{2poilat}
A= \frac{g^2_{d+1}N}{a^{d-1}} \frac{d-1}{N_s} \sum_{n=0}^{N_s-1} 
\int_{-\pi}^\pi \frac{d^dq}{(2\pi)^d} \left[\frac{2 \sin^2\frac{2\pi
      n}{N_s}}{D_n^2} - \frac{\cos\frac{2\pi n}{N_s}}{D_n} \right]~.
\end{equation}
We can combine the two terms in this equation by using a discrete
version of the integration by parts discussed in
Ref.~\cite{Kawai:1980ja}. First we need the (forward) derivative of
$D_n$
\begin{equation}
\label{Dpro}
D_{n+1}- D_n \equiv \nabla D_n = 2 \sin \frac{2\pi n}{N_s} \sin
\frac{2\pi}{N_s} + 4 \cos\frac{2\pi n}{N_s} \sin^2\frac{\pi}{N_s}~.
\end{equation}
By using this result, we can rewrite the first term in~\eqref{2poilat}
as follow
\begin{equation}
\label{tr1}
\sum_{n=0}^{N_s-1} \frac{2 \sin^2\frac{2\pi n}{N_s}}{D_n^2} =
\sum_{n=0}^{N_s-1} \frac{\sin\frac{2\pi n}{N_s}\; \nabla
  D_n}{\sin\frac{2\pi}{N_s} D_n^2} = 
  -\sum_{n=0}^{N_s-1}  \frac{\sin\frac{2\pi
      n}{N_s}}{\sin\frac{2\pi}{N_s}} \left[\nabla \frac{1}{D_n} +
  \frac{\nabla D_n}{D_n}\, \nabla\left(\frac{1}{D_n} \right)\right]~,
\end{equation}
where the term added in the second step vanishes due to the
periodicity of $D_n$. At the first order in the continuum limit
($N_s\to\infty$) the second term of this equation vanishes and we
obtain the relation used in Ref.~\cite{Kawai:1980ja}. It is easy to see
that the discrete analogue of an integration by parts involves the
backward derivative $\nabla^* g_n \equiv g_n-g_{n-1}$
\begin{equation}
  \label{eq:discrpart}
  \sum_n\left[\nabla f_n\right] g_n = -\sum_n f_n
  \left[\nabla^*g_n\right]~.
\end{equation}
By using this relation for the first term of Eq.~\eqref{2poilat} we
see that it cancels the second term, thus the total amplitudes becomes
\begin{equation}
  \label{sp}
  A = \frac{g^2_{d+1}N}{a^{d-1}} (d-1)\int_{-\pi}^\pi \frac{d^dq}{(2\pi)^d} 
  \left[ \frac{1}{N_s} \sum_{n=0}^{N_s-1}  \frac{\sin\frac{2\pi
        n}{N_s}}{\sin\frac{2\pi}{N_s}} \frac{\nabla D_n}{D_n}\,
    \nabla\left(\frac{1}{D_n}\right)  \right]~.
\end{equation}
It is possible to perform explicitly the sum over the discrete modes
of the momentum in the compact dimension. The idea is to rewrite the
sum as a contour integral; this can be done by promoting the the
combination $\ex{2\pi\ii/N_s}$ to a complex variable $z$. Then $D_n$
is substituted by the function
$D(z)=4\sum\sin^2\frac{q_i}{2}-(z+z^{-1}-2)$ and $D_{n+1}$ by
$D(\ex{\frac{2\ii\pi}{N_s}} z)$.  Then we multiply the complex
function obtained from~\eqref{sp} by the function $1/(z^{N_s}-1)$
which has poles at $z=\ex{2 \pi\ii n/N_s}$ for any integer $n$. Then
the square parenthesis in~\eqref{sp} is equal to
\begin{equation}
  \label{su1}
  \Big[\ldots\Big]= \frac{1}{2\pi\ii} \oint_C \frac{dz}{z}
  \frac{z-z^{-1}}{\ex{\frac{2\ii\pi}{N_s}}-\ex{-\frac{2\ii\pi}{N_s}}}
  \left(\frac{2}{D(z)} - \frac{1}{D(\ex{\frac{2\ii\pi}{N_s}} z)} -
    \frac{D(\ex{\frac{2\ii\pi}{N_s}} z)}{D^2(z)} \right)
  \frac{1}{z^{N_s}-1}~,
\end{equation}
where the contour $C$ is the union of an anticlockwise circle of
radius slightly bigger than one, and a clockwise circle with radius
slightly smaller than one. By supposing that the function in the
parenthesis of~\eqref{su1} does not contain additional poles on the
circle of unit radius (this is certainly the case for generic values
of $q_i$), one can apply Cauchy theorem and recover the sum in its
original form. Since the integrand is well behaved at infinity, we can
also deform the contours and sum all the residues whose modulus is
different from one. In this case the relevant poles are $z=0$, at
$z=\ex{\pm\tilde\phi}$ for the terms $1/D(z)$, and at
$z=\ex{\pm\tilde\phi}\,\ex{-\frac{2\pi\ii}{N_s}}$ for the term
containing $D(\ex{\frac{2\ii\pi}{N_s}} z)$, where
\begin{equation}
  \label{poles}
\tilde\phi = {\rm arccosh}\left(1+ 2 \sum_{i=1}^d
  \sin^2\frac{q_i}{2}\right)~.
\end{equation}
The residues of the poles of the first term (proportional to $1/D(z)$)
in~\eqref{su1} sum up to zero, while the remaining contributions
combine to yield a very simple expression
\begin{equation}
  \label{latre}
  A=\frac{g^2_{d+1}N}{a^{d-1}} (D-2) \left[ N_s \int_{-\pi}^\pi 
    \frac{d^dq}{(2\pi)^d} \frac{\ex{N_s
        \tilde\phi}}{\left(\ex{N_s\tilde\phi}-1\right)^2} \right]~.  
\end{equation}
Clearly, in the large $N_s$ limit, this integral is dominated by the
infrared region of low momenta; in fact when $q\to 0$ then also
$\tilde\phi\to 0$, while for physical momenta of the order $1/a$
(i.e. finite $q$) the integrand is exponentially suppressed. So, in
this limit, we can approximate the square parenthesis in~\eqref{latre}
as follow
\begin{equation}
  \label{larN}
  \Big[\ldots\Big] = N_s \int \frac{d^dq}{(2\pi)^d}
\frac{\ex{N_s q}}{\left(\ex{N_s q}-1\right)^2} = 
-\int \frac{dq}{(2\pi)^d} \Omega_{d-1} q^{d-1} 
\frac{d}{dq} \left(\frac{1}{\ex{N_s q}-1}\right)~,
\end{equation}
where $\Omega_{d-1} = 2\pi^{d/2}/\Gamma(d/2)$ is the volume of the
$d$-dimensional sphere of unit radius. Then we can integrate by parts
and use
\begin{equation}
  \label{Gz}
  \int_0^\infty dx \frac{x^{a-1}}{\ex{x}-1}=\Gamma(a) \zeta(a)~
\end{equation}
to obtain a compact formula for the 2-point function
\begin{equation}
  \label{mala1}
 \delta m^2 \sim \frac{D-2}{(2\pi)^d} \frac{g_{d+1}^2 N}{(2\pi R)^{d-1}}
  \frac{2\pi^{\frac{d}{2}}}{\Gamma\left(\frac{d}{2}\right)} \Gamma(d)
    \zeta(d-1)~.
\end{equation}
By using Legendre's duplication formula
\begin{equation}
  \label{legf}
  \Gamma(d) = \frac{2^{d-\frac{1}{2}}}{\sqrt{2\pi}} \Gamma\left(\frac
    d2\right) \Gamma\left(\frac{d+1}{2}\right)~,
\end{equation}
we can bring the lattice result to the same form found in the string
theory derivation of the previous section~\eqref{strft}
\begin{equation}
  \label{mala}
\delta m^2 \sim \frac{g_{d}^2N}{\pi^{\frac{d+1}{2}}}\frac{(D-2)}{(2\pi
R)^{d-2}}\zeta(d-1)
 \Gamma\left(\frac{d+1}{2}\right)~.
\end{equation}

\subsection{Wilson Fermions in the loop}

\begin{figure}[ht]
  \centering
  \begin{tabular}[h]{lc}
(a) fermion sunset: &
    \parbox{.9truein}{
      \epsfig{figure=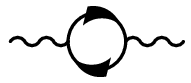}
      } \\
& \\
(b) fermion tadpole: &
     \parbox{.9truein}{
      \epsfig{figure=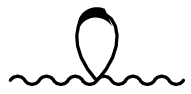}
      } \\
\end{tabular}
  \caption{Fermionic contributions to the gluon two--point function. }
  \label{fig:latferm}
\end{figure}

We expect that the same pattern seen in the previous section arises
for the loop contribution of any massless particle coupled in a way
that respects the higher dimensional gauge invariance. In this section
we show that this is indeed the case when considering Wilson fermions
minimally coupled to the higher dimensional gauge field. For the sake
of simplicity we will choose the Wilson parameter to be one
($r=1$). Of course, in the lattice Lagrangian for the Wilson fermions,
the chiral symmetry is broken and thus a mass term for these fermions
is generated through quantum corrections. In order to have a vanishing
effective mass one would need to add fine tune counterterms that
cancel these corrections. Since here we will focus only on the 1--loop
contribution to the scalar mass, the counterterms for the fermion mass
does not play any role and we will neglect this point. The fermion
contribution to the 1--loop function with two external scalars is
given by the diagrams in Fig.~\ref{fig:latferm}. Even if the lattice
Feynman rules for Wilson fermions are rather different from those of
the gluons, we see that the computation can be done by following
closely the same steps described in the previous sections. Again we
focus on the case of zero-momentum external particle, since we want to
extract the mass corrections from the 2-point function.

By using the Feynman rules listed in Ref.~\cite{Kaste:1997ks}, we obtain
for the first diagram in Fig.~\ref{fig:latferm}.a:
\begin{equation}
\label{sun}
A_a = -\frac{g^2_{d+1} T(F)}{a^{d-1}} \frac{c_D}{N_s}
\sum_{n=0}^{N_s-1} \int_{-\pi}^\pi \frac{d^dq}{(2\pi)^d} \left[ 
 \frac 12 \left(\sin\frac{4\pi n}{N_s} + D_n \sin\frac{2\pi
   n}{N_s}\right)^2 \frac{1}{D^f_n} -1\right] \frac{1}{D^f_n} ~,
\end{equation}
where $T(F)$ is the index of the fermion representation, $c_D$ counts
the physical polarizations of the fermion and $D_n^f$ is the Wilson
propagator 
\begin{equation}
\label{prof}
D_n^f = \sin^2\frac{2\pi n}{N_s} +  \sum_{i=1}^d \sin^2{q_i} +
        \frac{1}{4} D_n^2~.
\end{equation}
The contribution of the tadpole diagram (see Fig.~\ref{fig:latferm}.b) is
\begin{equation}
\label{tadp}
A_2 = -\frac{g^2_{d+1} T(F)}{a^{d-1}} \frac{c_D}{N_s} \sum_{n=0}^{N_s-1}
\int_{-\pi}^\pi \frac{d^dq}{(2\pi)^d} \frac{\sin^2\frac{2\pi n}{N_s} -
\frac 12 \cos\frac{2\pi n}{N_s} D_n}{D^f_n}~.
\end{equation}
By combining the two diagrams we obtain an expression that has a
structure similar to Eq.~\eqref{2poilat} 
\begin{equation}
\label{2pf}
A = -\frac{g^2_{d+1} T(F)}{a^{d-1}} \frac{c_D}{N_s} \sum_{n=0}^{N_s-1}
\int\limits_{-\pi}^\pi \frac{d^dq}{(2\pi)^d} \left[\frac 12
\left(\frac{\sin\frac{4\pi n}{N_s} + D_n \sin\frac{2\pi
n}{N_s}}{D^f_n}\right)^2 - \frac{\cos^2\frac{2\pi n}{N_s} + \frac 12
\cos\frac{2\pi n}{N_s} D_n}{D^f_n}\right].
\end{equation}
Then we will follow the same approach used in computing the bosonic
loop: we start by focusing on the first term and rewrite it in terms
of the variation of the fermionic propagator $D_{n+1}^f- D_n^f \equiv
\nabla D_n^f$ 
\begin{equation}
\label{ibp1}
\sum_{n=0}^{N_s-1}  \frac 12 \left(\frac{\sin\frac{4\pi n}{N_s} +
  D_n \sin\frac{2\pi n}{N_s}}{D^f_n}\right)^2 = 
  \frac{1}{2 \sin\frac{2\pi n}{N_s}} \sum_{n=0}^{N_s-1}
  \frac{\left(\sin\frac{4\pi n}{N_s} +  D_n \sin\frac{2\pi
  n}{N_s}\right) \nabla D^f_n }{(D^f_n)^2}~,
\end{equation}
where we have discarded all the terms that sum up to zero due to the
periodicities of the trigonometric functions. The we can
rewrite~\eqref{ibp1} as follow
\begin{equation} 
\label{ibp2}
  \sum_{n=0}^{N_s-1}  \frac{\sin\frac{4\pi n}{N_s} + D_n
  \sin\frac{2\pi n}{N_s}}{2 \sin \frac{2\pi n}{N_s}} \left(-\nabla
  \frac{1}{D^f_n} - \frac{\nabla D^f_n}{D^f_n} \nabla \frac{1}{D^f_n}
  \right)~.
\end{equation}
We can now use~\eqref{eq:discrpart} and ``integrate'' by parts the
first term in the parenthesis. In this way we see that it precisely
cancels the second term in~\eqref{2pf}. Thus the full 2-point
amplitude is
\begin{equation}
\label{2pf2}
A = -\frac{g^2_{d+1}T(F)}{a^{d-1}} \frac{c_D}{N_s} \sum_{n=0}^{N_s-1}
\int_{-\pi}^\pi \frac{d^dq}{(2\pi)^d} \frac{\sin\frac{4\pi n}{N_s} +
D_n \sin\frac{2\pi n}{N_s}}{2 \sin \frac{2\pi n}{N_s}} 
\left(\frac{2}{D^f_n} - \frac{1}{D^f_{n+1}} -
\frac{D^f_{n+1}}{(D_n^f)^2} \right)~.
\end{equation}
As in the previous section we can rewrite this sum as a contour
integral. Before doing this, it is convenient to rewrite the
propagator for the Wilson fermion in the following form
\begin{equation}
\label{Wprre}
D^f_n = \left(1+2\sum_{\mu=1}^d \sin^2 \frac{q_\mu}{2} \right)
\left[ 4\sin^2\frac{\pi n}{N_s} + \frac{\sum_{i=1}^d \sin^2{q_i} + 4
    \left(\sum_{i=1}^d \sin^2{q_i}\right)^2}{1+2\sum_{i=1}^d
    \sin^2{q_i}}\right]~.
\end{equation}
In this way the fermionic result~\eqref{2pf2} will take a form that is
very similar to the one encountered in the bosonic case. In particular
the first parenthesis combines with the other $sin$'s in~\eqref{2pf2}
and the contour integral we find has the same analytical structure
as~\eqref{su1} 
\begin{equation}
  \label{fe1}
 A= -\frac{g^2_{d+1}T(F)c_D}{a^{d-1}}
\int\limits_{-\pi}^\pi \frac{d^dq}{(2\pi)^d}
\oint\limits_C \frac{dz}{2\pi\ii z} 
  \frac{z-z^{-1}}{\ex{\frac{2\ii\pi}{N_s}}-\ex{-\frac{2\ii\pi}{N_s}}}
  \left(\frac{2}{\hat D(z)} - \frac{1}{\hat D(\ex{\frac{2\ii\pi}{N_s}} z)} -
    \frac{\hat D(\ex{\frac{2\ii\pi}{N_s}} z)}{\hat D^2(z)} \right)
  \frac{1}{z^{N_s}-1}~,
\end{equation}
where $1/\hat D(z)$ has poles at $z=\ex{\pm\phi^f}$ with
\begin{equation}
  \label{polesf}
\tilde\phi^f = {\rm arccosh}\left(1+ \frac 12 
\frac{\sum_{i=1}^d \sin^2{q_i} + 4
    \left(\sum_{i=1}^d \sin^2{q_i}\right)^2}{1+2\sum_{i=1}^d
    \sin^2{q_i}} \right)~.
\end{equation}
Thus we have now rewritten the 1--loop fermionic contribution in the
same form as encountered in the bosonic computation and the only
difference is in the explicit relation between the position of the
poles and the momenta $q$. Thus the 1--loop fermion contribution to the
scalar mass is
\begin{equation}
  \label{latref}
\delta m^2=-\frac{g^2_{d+1}T(F)c_D}{a^{d-1}} \left[ N_s \int_{-\pi}^\pi 
\frac{d^dq}{(2\pi)^d} \frac{\ex{N_s
\tilde\phi^f}}{\left(\ex{N_s\tilde\phi^f}-1\right)^2} \right]~.
\end{equation}
As we have already seen, in the large $N_s$ limit, only the low energy
momenta contribute significantly to this integral $q \sim 1/N_s$. Then
we can expand~\eqref{poles} for small $q$'s and we see that, for low
energy momenta, we have again $\phi^f \sim |q|$. Thus, in this limit,
the fermion contribution to the scalar mass reduces to the result
derived in the effective field theory~\cite{Cheng:2002iz}
\begin{equation}
  \label{mala1f}
 \delta m^2 \sim -\frac{c_D T(F)}{(2\pi)^d} \frac{g_{d+1}^2}{(2\pi R)^{d-1}}
  \frac{2\pi^{\frac{d}{2}}}{\Gamma\left(\frac{d}{2}\right)} \Gamma(d)
    \zeta(d-1)~.  
\end{equation}

\subsection{Accidental extended supersymmetry on the lattice ?}

Even if our lattice analysis has been restricted to perturbation
theory, the results of this section suggest the exciting possibility
of realizing SYM theories with extended supersymmetry on the lattice
in an accidental way. As is well known, lattice regularization
breaks most of the symmetries of the Poincar\'e group and all
supersymmetries that are present in the continuum version of the same
theory. However Poincar\'e symmetries arise automatically in the
continuum limit, because all relevant or marginal operators that could
violate them are prohibited by some of the symmetries that are present
in the theory at finite lattice spacing. A similar observation applies
also to the four dimensional ${\cal N}=1$ SYM
theory~\cite{Curci:1986sm,Kaplan:1983sk,Kaplan:1999jn}: if one adds to
the standard Yang--Mills theory a {\em chiral} fermion in the adjoint
representation, then no dangerous operator, such as a mass term for
the fermions, can be dynamically generated and, at low energies, one
automatically recovers a supersymmetric theory. Of course from a
lattice prospective the challenging aspect of this program is to
simulate dynamical chiral fermions.

In the case of extended supersymmetry an additional complication
arises: even pure SYM contains scalars (the complex scalar of the
vector multiplet). Once supersymmetry is broken by the lattice
regularization, one expects that a relevant mass term for these
scalars is dynamically generated and thus apparently there is no hope
to get a supersymmetric theory at low energies without fine
tuning~\cite{Kaplan:1999jn}. This is indeed the case if the scalars
are described by site variables in a four dimensional lattice. Various
approaches have been suggested to overcome this problem, such as
deconstruction, or the idea of realizing some of the supersymmetric
generators at finite lattice spacing, see
e.g. Refs.~\cite{Kaplan:2002wv,Sugino:2003yb,Catterall:2004np,D'Adda:2004jb,Takimi:2007nn},
a recent review with extensive bibliographic references can be found
in Ref.~\cite{Takimi:2007nn}.

The results of this section suggests a different possibility: one
can use the Kaluza-Klein reduction on the lattice to engineer four
dimensional SYM theories with extended supersymmetries from a higher
dimensional ${\cal N}=1$ theory. After all, also in the continuum field
theories, this is the easiest way to construct SYM theories with
extended supersymmetry~\cite{Brink:1976bc}. In this approach the
scalar fields are the internal components of the higher dimensional
gauge field and so are described by link variables in the compact
directions of a higher dimensional asymmetric lattice. Then, at
distances shorter than the compactification scale $R$, the scalar
fields and the gauge field are on the same footing and both are
constrained by the higher dimensional gauge invariance. Thus no
dangerous contribution to the mass of the scalar fields can come from
the high energy modes (i.e. modes with energies bigger than $1/R$). On
the contrary the quantum corrections to the scalar mass are purely due
to finite size effects and only modes with energies lower than $1/R$
can contribute. In the limit $a \ll R$, these modes are completely
blind to the effects of the lattice regulator and thus to the
supersymmetry breaking effects of the regularized theory.  This is
clearly visible in Eqs.~\eqref{latre} and~\eqref{latref}: when $1\ll
N_s$ the two expressions reduce to those obtained in the continuum
effective field theory and thus they cancel when expected. The overall
normalization in these results basically counts the number of bosonic
and fermionic degrees of freedom that can contribute to the mass
corrections. For instance, if we choose $D=6$ and two compact
dimensions, there is a fermion/boson degeneracy and we obtain ${\cal
N}=2$ SYM\footnote{The explicit expressions~\eqref{mala1}
and~\eqref{mala1f} are valid in the case of a single compact
dimensions $s=1$. For two compact dimensions one of the integrals
in~\eqref{latre} and~\eqref{latref} becomes a sum, however the
mechanism described here still applies: the leading order contribution
is independent of the lattice spacing and cancel between fermion and
boson loops when expected.}.

Thus there is hope to describe extended SYM theories on the lattice
without having to fine tune the scalar masses by using a higher
dimensional lattice with a different number of sites in the compact
and uncompact dimension.  For scales that are bigger than the
compactification radius, but smaller that the size of the
``uncompact'' dimensions ($N_s \ll x \ll L$) the lattice theory should
reduce to a standard four dimensional gauge theory. Of course an
obvious drawback of this approach is that simulations might be very
expensive when the number of dimensions of the lattice is big (for
instance, we would need 6 compact dimensions to simulate ${\cal N}=4$
SYM). Moreover there are several points that require further study in
order to see whether this proposal can be realized in a practical
way. A first obvious question is whether the pattern we described is
general or is just a peculiarity of the 1--loop perturbation
theory. There are actually some indications that this mechanism is
indeed general. The distinction between high energy modes, constrained
by the higher dimensional gauge invariance, and the low energy ones,
constrained by the tree-level supersymmetry, does not seem to rely on
the 1--loop approximation. So one would expect that the higher loop
radiative corrections to the scalar mass follow the same pattern and
the leading contribution in the large $R$ limit is independent of the
UV cutoff. Indications in this sense come from the string analysis,
where the 1--loop case is not special. A more fundamental question to
be addressed is to check whether this approach can be used to study
the strongly coupled regime of an $\mathcal N=2$ supersymmetric theory on the
lattice. Doubts in this respect were raised in Ref.~\cite{Kaplan:1999jn},
where it was noticed that, starting from a weakly-coupled
six-dimensional theory, the dynamically generated scale in four
dimensions $\Lambda_4$ is exponentially suppressed in the large $N_s$
limit. In this case, the $a$-dependent corrections to
Eqs.~\eqref{mala},~\eqref{mala1f} are likely to spoil the accidental
supersymmetry at the scale $\Lambda_4$. However a {\em strongly
  coupled} six-dimensional starting point is needed in order to take a
continuum limit of the lattice description which keeps the radius $R$
and the four dimensional coupling $g_4$ finite.  Thus the problem
mentioned above does not appear in the scaling limit that is relevant to
study of a fixed four dimensional supersymmetric physics.  Another
open question concerns the other types of fine tuning that might be
necessary in order to recover a theory with extended
supersymmetry. For instance, one would expect to fine tune the quartic
coupling among scalars and the Yukawa couplings so that they are all
related to the gauge coupling constant. The interplay between higher
dimensional gauge invariance and tree-level supersymmetry described
here should be helpful also to avoid the fine tuning of the couplings.

On a more practical side, one should worry about the subleading
corrections to Eqs.~\eqref{mala} and~\eqref{mala1}. These corrections
will certainly spoil the low energy supersymmetry and we can suppress
them only in the large $N_s$ limit which is of course computationally
very expensive. If these corrections are small for a moderate number
of lattice points in the compact dimensions, then this approach can be
really transformed into a practical tool for analyzing ${\cal N}=2$ or
even ${\cal N}=4$ SYM on the lattice. Some indications in this
direction come from~\cite{Kaste:1997ks}, where it is pointed out that
already for $N_s=8$ the lattice artifact effects are only of the order
of $2\%$.

\sect{Conclusions}
\label{sec:concl}

In this paper we studied the quantum corrections to the mass of the
internal components of the gauge field in Kaluza-Klein
compactifications. Within an effective field approach this problem was
analyzed in detail
in Refs.~\cite{Hosotani:1983xw,Hatanaka:1998yp,vonGersdorff:2002as,Cheng:2002iz,Ghilencea:2001ug,Alvarez:2006we,Alvarez:2006sf}. The
main feature of this result is that it depends only on the
compactification scale $R$ and is independent of the UV cutoff
$\Lambda$ necessary to define the higher dimensional gauge theory (of
course we assume $\Lambda\gg 1/R$). Even if this is the case, it is
natural to wonder whether an effective field theory approach is
reliable, since one is summing over the whole tower of Kaluza-Klein
states which at a certain point will have masses bigger that then UV
cutoff itself. In order to answer this question we studied the same
problem in two different UV finite theories: string theory and lattice
gauge theories. In the first case the UV cutoff is set by the string
length $\sqrt{\alpha'}$, while in the second case the same role is
played by the lattice spacing $a$; both these theories represent local
and gauge invariant UV completions of higher dimensional Yang--Mills
theories.

The interesting result is that, in the regime $R\gg \sqrt{\alpha'},a$,
both the lattice and the string computations reproduce exactly the
same result found in field theory, thus justifying a posteriori the
approach used in
Refs.~\cite{Hatanaka:1998yp,vonGersdorff:2002as,Cheng:2002iz}. This
analysis clarifies also the mechanism that protects the effective
field theory results from the contributions of the modes with an
energy of the order of the UV cutoff: since $R\gg \sqrt{\alpha'},a$,
these very energetic modes see all dimensions on the same footing and
the constraints of the higher dimensional gauge invariance should be
taken into account. Thus, if we want to compute radiative corrections
to terms that would violate the higher dimensional gauge invariance,
we do not really need to know the UV details of the string or lattice
theories. It is sufficient to know that these UV completions respect
locality and gauge invariance and this ensures that the leading order
contribution to these terms is completely captured by an effective
field theory approach. By carrying out the computation in the full UV
finite theory, we see explicitly that the suppression of the UV modes
is of the order of $\ex{-R/\sqrt{\alpha'}}$ (or $\ex{-R/a}$), while
the contribution of the low energy modes reproduces the expected
effective field theory result.

The string analysis can be relevant for phenomenological applications
in the context of models with large extra dimensions. In particular it
would be interesting to generalize our computation to the amplitudes
involving external states with a non-zero Kaluza-Klein charge. This
case has been discussed in detail from the field theory point of
view~\cite{Kaplan:1999jn,Bauman:2007rt,Bauman:2008rr}. The string
analysis can either support the picture emerging from the field theory
computations or maybe indicate subtleties due to the high--energy
modes. Of course it would be interesting to carry out the same
quantitative analysis in the case of tachyon free string theory. This
might be directly useful in in the string phenomenological scenarios
where the standard model is engineered on D-branes, which usually
contain non-chiral exotic matter fields.

In the context of lattice gauge theories the problem of the radiative
corrections to the Kaluza-Klein scalars is interesting because of its
connection with the possibility of obtaining an accidentally
supersymmetric theory at low energies. This is why we have considered
explicitly also the contribution of (Wilson) fermions. Even if
technically the computation is more involved than its bosonic
counterpart, we do not find any particular surprise and the pattern
described in section~\ref{sec:latt} arises. There is certainly the
need of more study to see whether this proposal can be turned into a
concrete approach to supersymmetry on the lattice. In general, we hope
that setups suggested by D-brane constructions and/or
compactifications can provide useful suggestions on how to realize
supersymmetric theories on the lattice also beyond the case of Super
Yang--Mills. Of course it would be very interesting to try and include
also chiral multiplets and construct a lattice realization of more
complicated supersymmetric theories.

\bigskip

\noindent
{\bf Acknowledgments.} This work is partially supported by the EC
Marie Curie Research Training Network MRTN-CT-2004-512194. LDD is
funded by an STFC Advanced Fellowship. LDD would like to thank
G.~Veneziano for useful comments at the early stages of this work. RR
would like to thank L. Magnea and S. Sciuto for enlightening
discussions. We would like to thank SUPA for funding the workshop on
"Lattice gauge theory beyond QCD" in Edinburgh.

\bibliographystyle{h-physrev3.bst}

\end{document}